# EVIDENCES FOR AND THE MODELS OF SELF-SIMILAR SKELETAL STRUCTURES IN FUSION DEVICES, SEVERE WEATHER PHENOMENA AND SPACE

A.B. Kukushkin and V.A. Rantsev-Kartinov

RRC "Kurchatov Institute",
Moscow, 123182, Russia

## ABSTRACT

This paper briefly reviews the (i) evidences for self-similar structures of a skeletal form (namely, tubules and cartwheels, and their simplest combinations), called the Universal Skeletal Structures (USS), observed in the range $10^{-5}$ cm - $10^{23}$ cm in the high-current electric discharges in various fusion devices, severe weather phenomena, and space, (ii) the models for interpreting the phenomenon of skeletal structures, including the hypothesis for a fractal condensed matter (FCM), assembled from nanotubular dust, and (iii) probable role of FCM, which might be responsible for the USS phenomenon, in tornado, ball lightning, and waterspout.

## 1.   INTRODUCTION

The present paper continues the series of the reports [1-3] on the progress that has been achieved since the last Symposium in our studies of the long-lived filamentary structures of a skeletal form in the electric discharges in various fusion devices, severe weather phenomena, and space. Here we give a very brief review of:

(i) the status of evidences for self-similar structures of a skeletal form (namely, tubules and cartwheels, and their simplest combinations), called the Universal Skeletal Structures (USS) [4(A)], observed in the range $10^{-5}$ cm - $10^{23}$ cm [4(B)] in

- high-current electric discharges in various fusion devices (tokamaks, including the dust deposits in tokamaks; Z-pinches, plasma foci; laser-produced plasmas) [4(C), 5(B)],
- severe weather phenomena (tornado, hailstones, lightning-born long-lived luminous objects) [4(B,D,E)],
- space (supernova remnants and some galaxies of similar form, "colliding galaxies", etc.) [4(B,F)];

(ii) the models for interpreting the phenomenon of skeletal structures:

- a fractal condensed matter (FCM), assembled from nanotubular dust [1,4(C)] and possessing an enhanced longevity in the ambient plasma due to the shielding of FCM by the EM waves [4(G)],
- various types of conventional (i.e. FCM-free) plasma filaments of electric current [6],
- strongly twisted magnetic flux ropes ("heteromacs" [4(H),7]) which require an enhanced internal magnetic coupling,
- aggregation of a fractal, assembled from nanoparticles, in a decaying/cooling plasma [8], or in a cold peripheral plasma [9];

(iii) probable role of FCM, which might be responsible for the USS phenomenon, in some severe weather phenomena:
- probable role of nanodust in the origin of tornado [4(D)],
- a hybrid of aerogel and plasma models of ball lightning [4(E)],
- hypothesis for the origin of waterspout [10].

The full survey report, presented at the 6-th Symposium, is available on the English home page of USS project (http://uni-skeletons.narod.ru/English-main.htm). This paper is appended with the post-Symposium results on numerical modeling of skeletal structuring [34].

## 2. THE STATUS OF EVIDENCES FOR SKELETAL STRUCTURING IN THE LABORATORY

### 2.1 The problem of diagnosing the USS in fusion devices

The status of evidences for self-similar structures of a skeletal form (namely, tubules and cartwheels, and their simplest combinations), called the Universal Skeletal Structures (USS) [4(A)], observed in the high-current electric discharges in various fusion devices (tokamaks, including the dust deposits in tokamaks; Z-pinches, plasma foci; laser-produced plasmas) [4(C),5(B)], may be qualified as "status quo". This can be illustrated for the case of the recognized leader of the fusion race, namely tokamak devices.

The presence of skeletal structures at initial stage of electric discharge in tokamaks was shown in the database from tokamak T-6 [2,11,12]. The visible light images were taken with the help of electronic optical converter (EOC) during electric breakdown stage of discharge (namely, ~300 $\mu$s before appearance of signal in the Rogowski coil which measures the discharge electric current). Also, analysis [11] of capabilities of the visible light imaging of tokamak plasmas revealed an important limitation imposed by the toroidal and poloidal rotation of the plasma column: the frame imaging, with 15 $\mu$s time exposure, by an EOC *before* appearance of the discharge current was sufficient to find the tubules and cartwheels of diameters about few-several centimeters. Similar imaging after appearance of the plasma current didn't allow the resolution of the structures because of fast rotation of the plasma column, both in toroidal and poloidal directions. However, imaging at quasi-stationary stage of discharge in another regime of EOC's operation, namely a streak camera regime with time resolution ~ 1 $\mu$s, revealed the structures similar to those seen at the «dark» stage in the framing regime. It was concluded that the time resolution as high as ~ 1 $\mu$s is needed for a rotating tokamak plasma. This conclusion has been supported by the results of the fast camera imaging (ultra-high speed CCD camera) used at tokamaks NSTX and Alcator C-Mod [13]. The presence of an exposure threshold, few centimeters long radial correlation of fluctuations of plasma emissivity, and bursty appearance of centimeter sized fluctuations in the camera view field [13], may be interpreted as implicit evidences for unexpected structuring. The origin of these phenomena could be understood from analyses of imaging with higher time and space resolution. This seems to be the feasible in the very recent time [14]: time resolution as high as 3 nsec has been achieved with intensifiers with spatial resolution ranging from 64 pixels (poloidally) x 64 pixels (radially) to (1024 x 1080) pixels at the shortest exposure times. However, the survey [15] reports on the ultra-high speed movies of the turbulent structure and motion of this light emission, which were made in both NSTX and C-Mod with a spatial resolution of (64 x 64) pixels only… One may expect getting much deeper insight into the fine structure of filaments and localized structures (``blobs'') with the help of (1024 x 1080) pixels imaging. In fact, such an imaging can provide fusion community with a diagnostic technique capable of implementing the missed diagnostic opportunities of streak camera method in tokamaks, suggested in [16] and based on the recognizability of skeletal structures.

Another important issue is the analysis of dust deposits. Recently the progress in studying the films deposited in the vacuum chamber of tokamak T-10 has been achieved via placing the scanning tunneling microscope *in situ* and analyzing the dynamics of films' growth [17]. An analysis of the available data obtained with this technique revealed the presence of few-several microns sized coaxial pairs of the rings of the same diameter, immersed in the amorphous component (cf. similar structures in the scanning electron microscope images of the films in [5]).

Redeposited hydrocarbon films on plasma facing elements in tokamaks accumulate hydrogen isotopes. Hydrocarbon thin films on metal mirrors and thick flakes with a high deuterium content, redeposited under deuterium (D)-plasma discharges inside the T-10 tokamak vacuum chamber, have been studied by means of the Fourier-transform infrared reflection, electron paramagnetic resonance, and photoluminescence spectroscopy [18,19]. The films from T-10 were found to significantly differ from conventional, gas deposition-formed films by the presence of vibration modes of $CD_2$, $CD_3$ (carbon-deuterium) and $CH_{1-3}$, adjoined to $sp^3$ bonds in graphene. These results suggest these films to possess a strong fraction of "free" graphene sheets which, as known, are the raw material for production of carbon nanotubes. One more example of evidences for the capability of carbonaceous materials to produce unusual, skeletal structures under strong electrodynamic processing is the production of large carbon-based toroids (CBTs) under two-step laser irradiation of a mixed fullerene target [20]. Transmission electron microscopy clearly showed CBTs of diameter 0.2–0.3 μm with tubular diameters of 50–100 nm, but toroids as wide as 0.5 μm were observed. These structures are the small-scale relatives of skeletal structures revealed in [3,4(C)] in the images of laser-produced plasma (laser plume).

### 2.2 USS in severe weather phenomena

Major progress in identification of USS in severe weather phenomena (tornado, hailstones, lightning-born long-lived luminous objects) was achieved via analyzing the presence of skeletal structures in tornado [4(D,E)] in extension of the first results reported in [4(B)]. Main features of skeletal structures found -- with the help of the method of multilevel dynamical contrasting (MDC) of the images [4(H)] -- in the available databases of tornado's images, are as follows.

Skeletal structures are present in the main body of tornado and its close vicinity. General layout of skeletal structures in the main body of tornado is similar to that in a straight Z-pinch. The cartwheels on the axle-tree outside the main body of tornado are directed transversely to the main body – similarly to cartwheels seen at the periphery of hot plasma column of the Z-pinch. The main body of tornado possesses large tubular structures which are also directed transversely to the funnel (Fig. 1).

The resolvability of skeletal structures in the photos taken with large enough exposure suggests that these structures are moving slowly as compared to motion of gaseous component and therefore are, to a large extent, decoupled from the motion of this component.

An electric torch-like structuring (the shining butt-end of a truncated straight filament) of tornado's funnel is similar to that found in a very broad range of length scales [4(B)].

As far as the low-precipitation supercell are known to produce an intense hailstorm with large enough hailstones, the hailstones may say more about tornado than it was thought. It appeared difficult to identify the trend toward self-similarity in the available images of tornadoes. However, this trend is found to be quite distinct in the hailstones [4(D)].

A.B. Kukushkin and V.A. Rantsev-Kartinov

The existence of areas of highest tornado activity (first of all, Tornado Alley in Kansas, USA) may be associated with the delivery of nanodust material from volcanoes in Africa, according to identified statistics of atmospheric transport.

The evidences for the presence of skeletal structuring on the ocean surface during its strong perturbations by the hurricanes are given in [22].

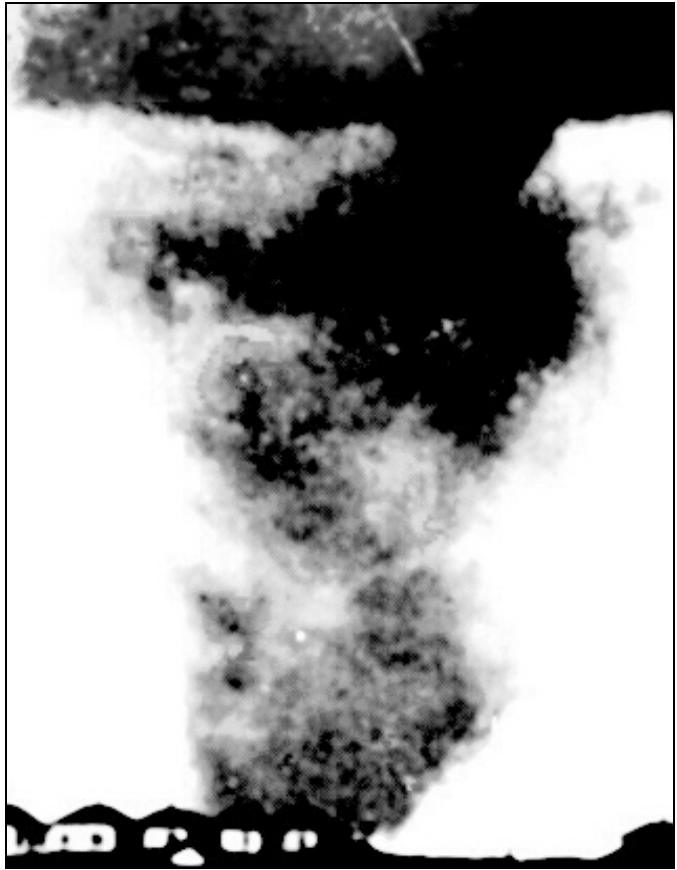

Fig. 1. The MDC-processed image of tornado at Norton, Kansas (June 1909), taken from collection [21], http://www.photolib.noaa.gov/historic/nws/images/big/wea00248.jpg.. The tubular formations in the main body of tornado are directed nearly horizontally and sometimes possess a resolvable coaxial structure.

### 2.3 USS in space

The evidences [4(B,F)] of skeletal structures in space (supernova remnants and some galaxies of similar form) may be appended with the evidences [23] which support the possibility to reinterpret the phenomenon of the so called "colliding galaxies" in terms of the hypothesis for a cold baryonic skeleton of the Universe [4(B,F),24]. According to this hypothesis, one can see such a skeleton in some critical, burning points (which are either the "sparkling" caused by the local disruption of the skeleton, or the shining open butt-end of a dendritic circuit). The most convincing example of such a picture seems to be a pair of the shining butt-ends of the broken straight filament. Such "colliding" galaxies have to be of the same size and possess the complementary picture of the butt-end, which characterizes the inhomogeneity of the break within the cross-section of the break.

Here we give an example, taken from [23], of reinterpreting the phenomenon of "colliding galaxies" (Fig. 2).

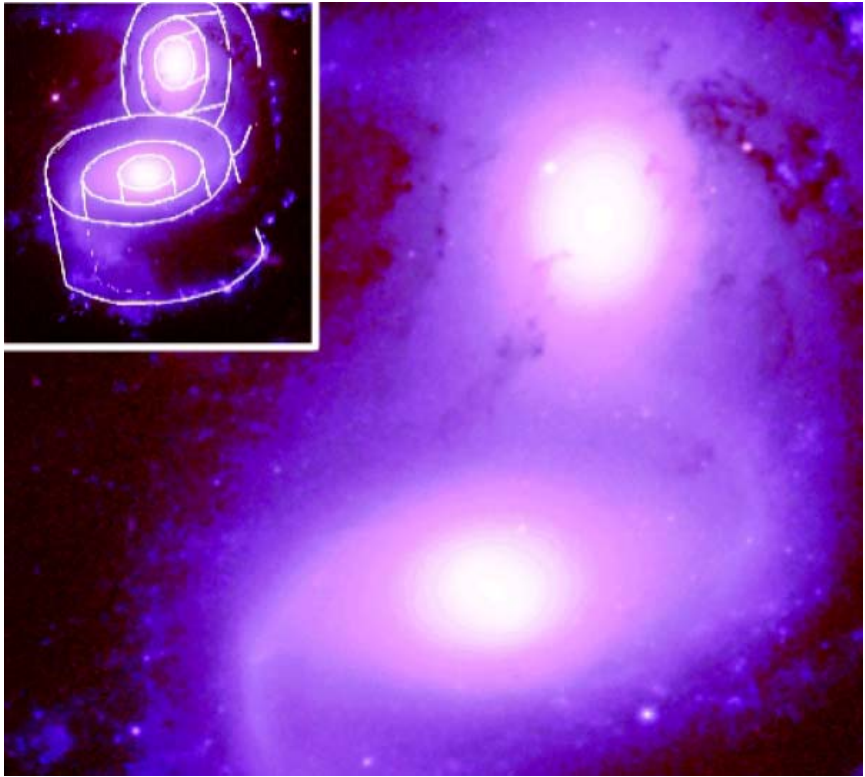

Fig. 2. Interacting galaxies NGC 7319A and NGC 7319B in the constellation Pegasus (original image is taken from the Hubble Space Telescope database [26] and processed [23] with MDC method). Diameter of these galaxies is ~ 3 $10^{22}$ cm. The butt-ends of the broken parts of hypothetical very cold (and invisible) filament are directed nearly perpendicularly. A schematic drawing of the reconstructed structuring is given in the insert.

More data on the filamentary structuring at cosmological length scales, which is derived from analysing the galaxies redshift surveys -- in extension of analysis [4(F)], are presented in [25].

## 3. THE MODELS FOR INTERPRETING THE PHENOMENON OF SKELETAL STRUCTURES

### 3.1 Conventional plasma filaments of electric current

Conventional approach to describing the filaments in plasmas is anyway based on Bennet-type equilibrium which requires the counterbalance of plasma pressure and the pinchning of electric current. The attempt to interpret, in such a frame, the skeletal structures, observed [4(C,H)] in tokamaks, Z-pinches and plasma foci, was undertaken in [6]. Possible configurations of current filaments in Z-pinch and tokamak plasmas were analyzed. First, it was suggested that a thin current-carrying beam injected in a plasma should be surrounded by a halo of countercurrents, in which case the resulting configuration may resemble a tubular structure. It was claimed that that skeletal structures [4(C,H)] (tubules, squirrel cage structure, and hexagons as a basic elementary block of tubules) can also be attributed to the fundamental mode of the conventional magnetic filamentation in the form of a "hexagonal parquet." Second, an analytic study of the phenomena governing the pattern of plasma structures, namely, tearing filamentation, two types of longitudinal beam bunching, and self-organization of the filaments, was carried out.

On the whole, the results [6] illustrate the opportunities of conventional MHD approaches. On the one hand, the conventional mechanisms of filamentation of electric current in plasmas do give a reasonable general picture of filament's formation and equilibrium. On the other hand, application to particular conditions at which the long-lived skeletal-type filaments are observed, reveals this approach to be still insufficient. For instance, the filaments directed across the strong magnetic field in tokamaks [4(C,H)], if interpreted as a purely plasma filament of electric current [6] would require the values of in-filament current (and of the above-mentioned halo countercurrent) at least comparable with the total discharge electric current. The latter currents would produce the Joule heating of plasma at a much different rate as compared with that inferred from the experiments.

### 3.2 Strongly twisted magnetic flux ropes

The first step in interpreting the unusual structuring [1,4(H)] revealed in the database from the former experiments in the Z-pinch facility – namely, waiving of the filaments of plasma emissivity, and the tree-like structure around main body of the Z-pinch plasma column, especially the presence of strata directed nearly transversely to the Z-pinch plasma column – was suggested partly by the physics of magnetic flux ropes in which the nearly force-free magnetic configuration was thought to be capable of self-sustaining (e.g., magnetic flux ropes in space), as well as by the success in confining the nearly force-free magnetic configuration in an electrically conducting chamber (spheromaks and reversed field pinches). The assumption of a strong confinement of magnetic field in the flux tubes, enabled us to treat the long-lived filament in plasmas as an elastic thread which, as everybody can immediately check via twisting the ordinary thread, can produce a compact, twisted loop directed transversely to the thread. In plasma, this would correspond to formation of an almost-closed helical heterogeneous magnetoplasma configuration (we called this configuration a *heteromac* [1,4(H)]). Such a branching-off process makes single filament a fractal, dendritic structure (Fig. 3).

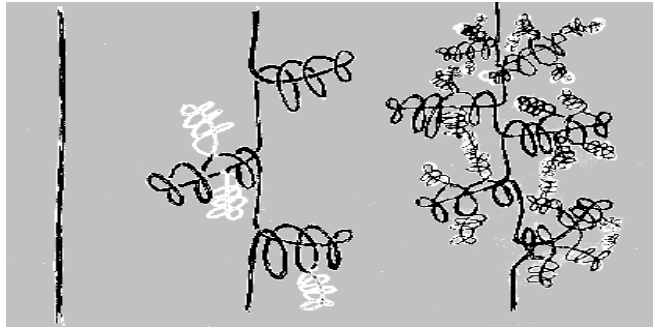

Fig. 3. A schematic drawing [1,4(H)] of successive branching of an originally one-dimensional filament (left drawing), which produces the heteromac(s) (center) and makes individual filament a fractal formation (right).

The heteromacs were suggested [7] to produce cellular, and bubble-like clusters. In particular, this hypothesis was applied to interpreting the Aurora phenomena in terms of a high-current Z-pinch. The discovery [7] that objects from the Neolithic or Early Bronze Age carry patterns associated with high-current Z-pinches provides a possible insight into the origin and meaning of these ancient symbols (primarily, petroglyphs). This conclusion is based on direct comparison of the graphical and radiation data from high-current z-pinches to these patterns.

The heteromac hypothesis still leaves open the question about the origin of the longevity of observed filamentation. Here, major difficulty is the unknown mechanisms for suppressing all the short-scale instabilities, which the plasma is rich with: the mater is that in the successful MHD simulations of filament's dynamics these instabilities are ignored. For instance, the two-fluid description of plasma -- with allowance for the Hall effect-produced "loosening" of the boundary between the plasma column and the ambient magnetic field -- makes this boundary pretty unstable in contrast to observations [2,4(H)] of long-lived filaments in the high-current electric discharges. The latter implies that successful MHD simulations may implicitly exploit the physics hidden at the length scales which are smaller than those treated in the codes.

### 3.3 Aggregation of a fractal, assembled from nanoparticles

One of the ways to interpret the networking of filaments in electric discharges has been proposed in the papers [8]. Post-discharge observations [27] of the fractal aggregates fastened to the walls of vacuum chamber enabled the authors [8] to suggest the observable "ball lightning" (BL) objects to possess a network of the aerogel type (which was called a skeleton in [8(A)]). The fractal skeleton was suggested to be assembled from nanoparticles produced by the plasma coagulation/clustering at the stage of plasma decay/cooling. This allowed [8(A)] the explanation of the lifetime of typical BL and of some other features of low and moderate energy BLs -- namely, of their elasticity and luminosity (in the model [8(A)], the BL's luminosity is due to hot-spot chemical combustion of skeleton).

The phenomenon of aggregation of a fractal, assembled from nanoparticles, in a decaying/cooling plasma or in a cold peripheral plasma may be extended even to short-lived plasmas, e.g. those produced in a laser plume (see, e.g., the survey [9]). In particular, analysis [9] of scanning electron microscope images of deposits on the glass collector, as formed via laser evaporation of iron target, showed the presence of the chains, which consist of nanowire sections and balls, rather than the presence of merely separate compact particles/blocks. Significantly, the authors [9] consider the induced dipole moment of nano-sized (~10 nm) macromolecules to be able to produce the chains not only for ferromagnetic materials.

### 3.4 Fractal condensed matter (FCM), assembled from nanotubular dust

The concept of a fractal condensed matter (FCM), assembled from nanotubular dust [1,4(C)] and possessing an enhanced longevity in the ambient plasma due to the shielding of FCM by the EM waves [4(G)], is based on appealing to exceptional electrodynamic and mechanical properties of their hypothetical building blocks. The prediction [1,4(C)] of the phenomenon of fractal skeletons in a wide range of lengths, starting from nanoscale structures, assumed, first of all, the following features of hypothetical building blocks: (i) enhanced cold emission of electrons by these blocks to facilitate the electric breakdown in laboratory discharges; (ii) enhanced ability to self-assemble macrosopic structures to form the skeletons of macroscopic size; (iii) enhanced mechanical strength of these blocks to assure the integrity and longevity of macroskeletons; (iv) most economic consumption of the relevant available material to produce stable long-range quantum bonds. These properties allow the build up of the Eiffel Tower-like, tracery structures (cf. Fig. 4) with minimal mass for a given strength of the entire construction.

The carbon nanotube (CNT), or similar nanostructures of other chemical elements, were suggested to be such blocks. Among the properties of CNTs, which that time were already identified in the literature and stimulated the choice [1,4(C)], we have to mention, first of all, exceptionally high mechanical strength, high enough electrical and thermal conductivity of individual nanotubes in a wide range of major parameters (that promised making them an ideal 1-D quantum wires), ability of CNTs to form large, macroscopic clusters.

A.B. Kukushkin and V.A. Rantsev-Kartinov

The self-assembling of skeletons was suggested to be based dominantly on *magnetic* phenomena. The indications on the ability of CNTs, and/or their assemblies, to trap and almost dissipationlessly hold the magnetic flux come from the following observations. Superconductor-like diamagnetism in the assemblies of CNTs at high enough temperatures was reported in [28,29]. The evidences were obtained for the self-assemblies of CNTs (which contain, in particular, the ring-shaped structures of few tens of microns in diameter) inside *non-processed* fragments of cathode deposits, at room temperatures, [28] and for the artificial assemblies, at 400 K [29]. However, observations of room-temperature superconductor-like diamagnetism in CNT assemblies are still limited to these two groups. The evidences and arguments for the room-temperature superconductivity in individual CNT, and in artificial and natural assemblies of CNTs, are summarized in [30].

We also have to mention observations of unexpected ferromagnetism of a *pure* carbon, namely that in the rhombohedral $C_{60}$, with a Curie temperature near 500 K [31], and room-temperature ferromagnetic nanotubes controlled by electron or hole doping [32]. The recent survey of experimental evidences for, and theoretical models of, the unexpected magnetism of carbon foams and heterostructured nanotubes is given in [33].

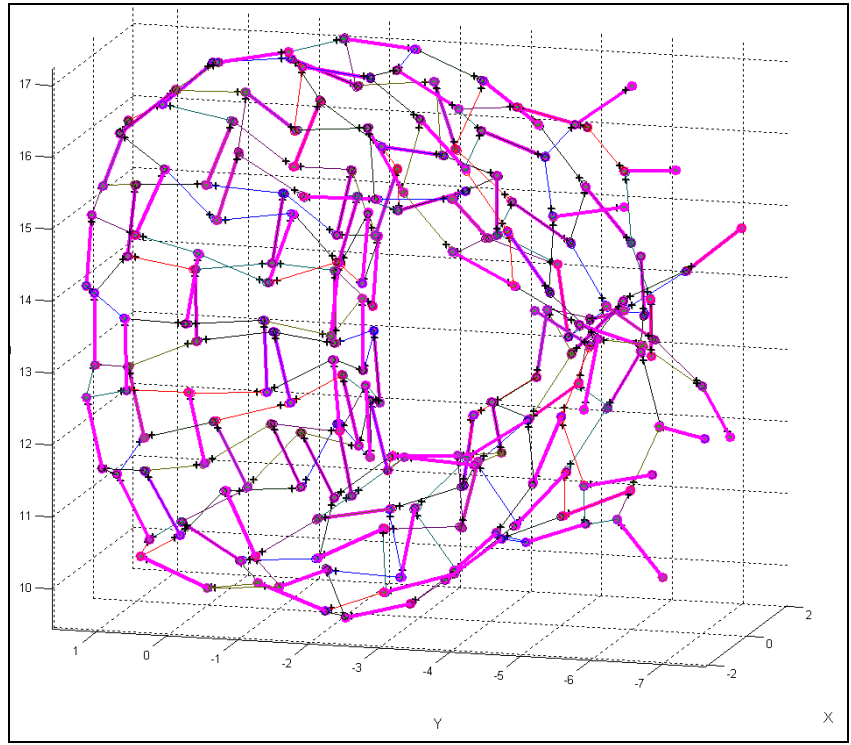

Fig. 4. The image of the results of numerical modeling [34] of the impact of a distant pulsed electric current on the tubular, originally straight skeletal structure, composed of 294 one-dimensional magnetic dipoles which sustain circular electric current in the wall of this skeleton (external current was passed along X-direction, the line of current is located in the point {Y= -15, Z= 15}), for more detail see [34]).

Despite the above experimental evidences and theoretical models need much stronger tests and confirmations, they justify explicit demonstration of the capability of magnetized nanotubular blocks to self-assemble the tubules of higher generations [1] and sustain the integrity of the assembled skeleton. An analysis of the capability of magnetized one-dimensional rods to sustain the integrity of the hypothetically formed tubular skeleton (i.e. the tubule of the 2-nd generation) was carried out in [34] in the frame of a simple electrodynamic model suggested there for describing such a skeletal matter. The capability of the model was illustrated with the example of how a straight tubular skeleton, which is composed of 294 dipoles and carry circular electric current in its wall, may be wrapped up by a distant pulsed electric current to make a toroid-like structure (Fig. 4). The results [34] may be interpreted as an illustration of the possibility of such skeletons -- if formed in the high-current electric discharges or under similar conditions - to form the torodal-like and cartwheel-like structures (cf. laser-induced production of large carbon-based toroids reported in [20], see the Q-shaped toroids in Fig. 3 of this paper).

## 4. PROBABLE ROLE OF FRACTAL CONDENCED MATTER IN SOME SEVERE WEATHER PHENOMENA

Probable applications -- both inside and outside the fusion science -- of the fractal condensed matter (FCM) which is assembled from nanotubes and might be responsible for the USS phenomenon have already been briefly discussed in [3]. These included facilitation of electric breakdown of the working gas in the discharge chamber, control of the nonlocal, non-diffusional component of heat transport in magnetically confined fusion plasmas, production of a new type of nanomaterial [5], reconsideration of the "dark matter" problem in astrophysics and cosmology [4(F)]. Here we discuss the following problems: the probable role of nanodust in the origin of tornado [4(D)], a hybrid of aerogel and plasma models of ball lightning [4(E)], and hypothesis for the origin of waterspout [10].

### 4.1 Probable role of nanodust in the origin of tornado

The current status of treating the tornado's mystery may be illustrated with the recent findings [35] from the project VORTEX (Verification of the Origins of Rotation in Tornadoes Experiment): "VORTEX has produced a number of troubling new findings. For example, it appears that perhaps *many fewer* supercells and mesocyclones produce tornadoes than scientists originally believed… Further, we have learned that the *difference between tornadic and non-tornadic mesocyclones can be very, very subtle*…" The conclusions [35] well justify a search for the contribution of an unrecognized mechanism to initiation of tornadoes. Below we present major points of the hypothesis [4(D)] for the possible contribution of nanodust (or generally speaking, of a FCM component) to the origin of tornadoes.

Tornadic thundercloud may possess an *internal skeleton* which is hidden in the water vapor, air-based dust, etc. The skeleton may be composed of the *nanodust*. The nanodust particles (presumably, carbon nanotubes or similar nanostructures containing other chemical elements) in the skeleton may, as it was suggested [1,4(C)], be *magnetically* coupled by the magnetic flux trapped in the nanotubular block. Such a skeleton is flexible and restructurable, being a sort of the *fractal aerogel*. The fractality of the skeleton is suggested by the analysis [4(D)] of available observations (see Sec. 2 in [4(D)])

The skeleton in tornadic thundercloud may be responsible for the *fast long-range transport of electricity* (e.g., with respect to electric charge acquired by the skeleton during condensation of charged water drops on it) and for accumulation of large enough electric charge in certain points determined by the geometry/morphology of the skeleton (e.g., at the skeleton's edges).

This implies that actually the difference between tornadic and non-tornadic thunderclouds is determined by the transport properties (in particular, by the very presence) of an internal skeleton, namely by the ability of the skeleton to collect, transport and focus the electric (and magnetic) energy. In particular, skeleton as a condensation center may substantially speed up the conversion of the latent heat into gas/plasma motion.

The localization of large enough electric charge on/around the skeleton initiates a sort of electric breakdown between the thundercloud and the Earth. Thus, the initial phase of tornado may be treated as an electric breakdown process which is eye-visible in the real time. Presumably, most frequently the cathode is the thundercloud, the virtual anode is on the Earth surface. Thus, tornado's initiation is suggested to be an electrostatic instability caused exclusively by the presence of nanodust and its special, skeletal structuring.

Tornado's column/funnel may be interpreted as a *long-lived filament* of electric current which is being resulted from the process of aforementioned electric breakdown (the value of electric current may be still weak for pinching the plasma in the tornado column). The skeleton of this filament is being formed by the restructured blocks of the skeleton in the thundercloud during a «pullout» of the parent thundercloud's skeleton downwards by the electric force which disturbs the balance of the forces in the stable state of thundercloud. (Here, longevity implies that the lifetime largely exceeds that of the lightning).

Rotation of tornado's column/funnel seems to be a consequence of local restructuring of the skeletal component inside the thundercloud. That's why the rotation in the tornadic mesocyclone is actually ***not*** a source of tornado's initiation (cf. [35]). Therefore, the problem of tornado rotation may be decoupled from the problem of tornado's initiation and has to be treated in terms of the behavior of the ambient gaseous and aerosol components of the thundercloud in the presence of electric (and magnetic) field provided by the skeleton in the course of its restructuring.

The *early diagnostics* of tornado has to be aimed at identifying a skeleton inside a thundercloud and especially at those patterns of skeleton's behavior/restructuring which might be dangerous for tornado's initiation.

## 4.2 A hybrid of aerogel and plasma models of ball lightning: an inductive storage wildly formed by a nanotube-assembled skeleton

An analysis of the probable role of nanodust in severe weather phenomena encourages us, in extension of [4(I)], to come back to the ball lightning (BL), another atmospheric phenomenon whose longevity is anomalous from viewpoint of plasma chemistry kinetics. A fusion of the following two hypotheses, namely: (1) for the probable role of nanotubular dust in the observed anomalous longevity of filamentary structures in electric discharges [4(E)], and (2) for the origin of BLs, especially of those of *high energy store*, treated as a rare atmospheric analog of the much more investigated phenomenon - the filaments of electric current in high-current laboratory discharges, enabled us to suggest a unified treatment of both phenomena that requested, though, substantial modification of the existing approaches to each of them. Regarding the BL, this resulted in the following alterations of the concept [8] of the presence, in BL, of a rigid skeleton of aerogel type.

(i) A (ball-shaped) skeleton is electrodynamically assembled from solid nanoblocks *in advance* of plasma formation (contrary to formation of nanoblocks at the stage of plasma decay/cooling by the plasma coagulation/clustering and further assembling of a fractal from such blocks [8]). Thus, we suggest to interpret post-discharge observations [27] of fractal aggregates fastened to the walls of the vacuum chamber as being caused by the nucleation/deposition of the vapor at a skeleton formed yet at the electric breakdown stage of discharge.

(ii) A search for nanoblocks which may *facilitate* electric breakdown (e.g., via anomalous emission of electrons by these blocks, both via thermal and electric field-emission mechanisms), assemble *macroscopic* skeleton, and trap and hold, with *low dissipation*, the magnetic field and high-frequency (HF) EM field, suggested a new candidate - nanotubular structures (most probably, carbon nanotube (CNT)).

(iii) A substantial store of *magnetic* (and electromagnetic) energy in BL is possible (~ > 1 MJ), which is finally responsible for luminosity of BL during anomalous long time (contrary to luminosity due to hot-spot *chemical* combustion of skeleton [8(A)]), and for elasticity of BL.

The above approach to BL gives the following hybrid of the «aerogel» model [8] and the «plasma» models of BL (for the comparison of these models see survey [8(A)]).

In the plasma models, BL is considered to be a product of electric discharge which gives the spheromak-like magnetic configuration. This assumption is well supported by the experimental data on spheromak production and confinement in a magnetic flux conserver, and by the success of theoretical prediction [36] of evolution towards force-free magnetic configuration. The strong point of plasma models of BL is that this approach relies on the well-known advantages of spheromak configuration for confining the plasma. The weak point of any plasma model is that the lifetime, $(\tau_E)_{pl}$, of plasma thermal energy is too short – at least, because of radiative losses. Indeed, available experimental data on $\tau_E$ in laboratory plasmas suggest that, for energy store of 10 kJ (which is typical value both for the laboratory spheromaks, without auxiliary heating, and for BLs ), the value $(\tau_E)_{pl} \sim$ 1-10 ms is much smaller than typical BL's energy lifetime, $(\tau_E)_{BL} \sim$ 10 s. It looks like the spheromak-like magnetic configuration may actually result from a localized electric discharge in the atmosphere (e.g. from lightning stroke) but the plasma itself is not able to confine the stored magnetic field for long. Therefore, one has to append the plasma model with a highly conducting matter capable of confining the residual magnetic field. Such a matter has to cope with the heritage of the initial, «plasma» stage of BL, which, for BL of diameter $D_{BL}$ and of stored magnetic energy $E_{MF}$, may be estimated as follows:

$$I_p(MA) \sim \sqrt{\left[\frac{E_{MF}}{10\,kJ}\right]\left[\frac{10\,cm}{D_{BL}}\right]}, \qquad (1)$$

$$B_p(T) \sim B_t(T) \sim 3\sqrt{\left[\frac{E_{MF}}{10\,kJ}\right]\left[\frac{10\,cm}{D_{BL}}\right]^3} \qquad (2)$$

where $I_p$ is the total electric current (in MegaAmpere units), $B_p$ and $B_t$ are the space-averaged poloidal and toroidal magnetic fields (in Tesla units), respectively, and

$$p_{MF}(MPa) \sim 10\,(E_{MF}/10\,kJ)(D_{BL}/10\,cm)^{-3} \qquad (3)$$

is magnetic pressure (in MegaPascal units) which for typical BL is as high as ~100 atmospheres.

Let us assume that a BL's skeleton is formed according to items (i) and (ii) of this Section. Note that survivability of skeleton at plasma stage may be possible due to «wild cable» mechanism [1(G)]: an intense luminosity at this stage comes from plasma which is intermittently isolated from the *cold* skeleton by the Miller's force of HF EM field. The total mass of CNT skeletal matter, $M_{CNT}$, which may carry the current of Eq. (1), may be estimated in terms of the typical diameter of CNT, $d_{CNT}$; the average number of walls in the CNT, $N_W$; and the value of electric current through *individual* CNT, $I_{CNT}$:

$$M_{CNT}(g) \sim 0.1\sqrt{\left[\frac{E_{MF}}{10\,kJ}\right]\left[\frac{D_{BL}}{10\,cm}\right]\left[\frac{10\,\mu A}{I_{CNT}}\right]\left[\frac{d_{CNT}}{nm}\right]}N_W. \qquad (4)$$

Note that the current density of $10^7$-$10^9$ A/cm$^2$ through individual CNT is possible, with $I_{CNT}$ being as high as ~10-1000 µA. It follows from Eq. (4) that for typical BL the skeleton's specific weight is less than that of the air. Assuming that the BL's energy lifetime $(\tau_E)_{BL}$ is determined by the dissipation of magnetic field through Joule heating of

A.B. Kukushkin and V.A. Rantsev-Kartinov

CNT skeletal matter, one may estimate the upper value of the resistivity, $\rho_{CNT}$, of this matter:

$$\rho_{CNT}(\Omega cm) \leq 10^{-8} \left[\frac{D_{BL}}{10\,cm}\right]^2 \left[\frac{10\,s}{\tau_{BL}}\right]. \qquad (5)$$

Such a hybrid of «aerogel» and «plasma» models of BL is aimed at eliminating the weak points of these approaches via appealing to the properties of CNT's which, however, require much stronger confirmation.

The verification of the above approach may include a comparison of the fine structuring found in the long-lived skeletal structures in the high current laboratory discharges with that of (i) aerogels and (ii) ball-shaped luminous objects, not necessarily BL's (!), produced by the powerful lightning. Such a comparison (Figs. 5 and 6) supports the above approach.

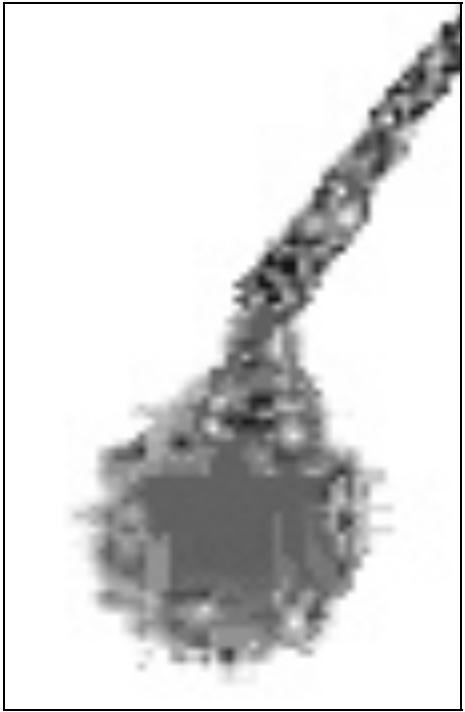

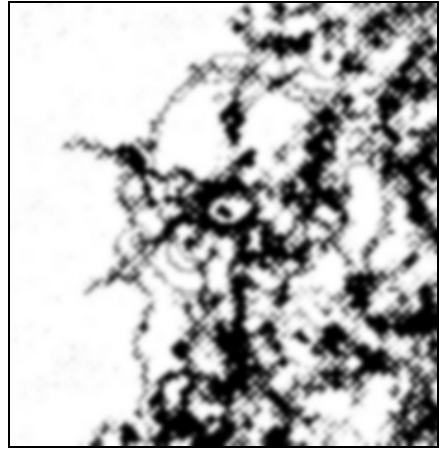

**Fig. 5.** A fragment of the image of the end point of the powerful lightning which took the form of a ball-shaped structure of the estimated diameter of several meters (the original image is taken from the web-site of Long Island Lighting Co.). The processing of the image with the method of multilevel dynamical contrasting [4(H)]) reveals, on the right-hand side of the ball, the elliptic image of the cartwheel-like structure with the cartwheel's axis directed perpendicular to electric current in the lightning. It looks like the ball may contain a tubular structure (seen as a thick black band) whose axis is directed transversely to lightning's direction.

**Fig. 6.** A fragment of scanning electron microscope image [37] of the aerogel thread obtained in the experiments on laser irradiation (of intensity ~ $10^7$ W/cm$^2$ ) of metallic target in external electric field (image width is ~ 8.5 μm). Such an aerogel is found [37] to be composed of ~10 nm thick filaments. A cartwheel-like structure in the center of the image is located on the left border of the thread's image, where it is easier to identify such a structure. Typical location of cartwheels at thread's surface (namely, perpendicularly to thread's axis) is similar to location of cartwheels identified [2,4(C)] in the high-resolution images of Z-pinch plasma column.

### 4.3 Waterspout as a special type of atmospheric aerosol dusty plasma

The problem of waterspout formation was considered in [10]. It was based on revelation of the probable presence of skeletal structures in the ocean [22].

Major hypotheses for formation of skeletal structures on the ocean surface were stimulated by the analysis of databases on hurricane trajectories in Atlantic Ocean for about hundred years period. This analysis suggested that Africa, with its volcanoes and deserts, may be responsible for formation of big structured clouds, which usually accompany the powerful severe weather phenomena in this ocean. Similar analysis of typhoon trajectories led to a conclusion on the probable responsibility of volcanoes in Japan and Oceania. Indeed, volcanoes may deliver a structure-forming dust to the atmosphere. Under the action of the Earth electric field and atmospheric electricity such a dust may form the skeletal structures (SS) in the clouds [4(D)], which may ultimately be deposited on the ocean surface [22]. Thanks to active surface of SS, the gas, dissolved in the water, may be absorbed at SS surface thus providing SS with some buoyancy. In the rough sea water, these structures may be partly destroyed and their fragments may compactly fill in the space in the above-mentioned coaxial tubular structures. With the coexistence of three phase states (namely, solid, liquid and gas) on the SS's surface, the presence of SS provides the actions of surface tension force even under the water level and, consequently, may result in the strengthening and sticking together of separate blocks into a unified skeletal structure.

The presence of such structures on the ocean surface in the vertically aligned position may significantly facilitate the electric breakdown between the ocean surface and the charged cloud [10], thanks to intense production of a highly disperse charged water aerosol. Under the action of electric field the aerosol lifts up at a high speed and involves the neutral gas in this motion. The drop of atmospheric pressure in such an air column leads to suction of the ambient gas by the entire surface of the column. As far as the air contains weakly ionized plasma, its radial motion in the presence of vertical component of the Earth's magnetic field, and of uncompensated electric charge, may induce the rotation which can accelerate the gas in the column and its surroundings to high rotation velocities. The column itself and its far wings rotate in opposite direction [10].

The waterspout appears to be an electrostatic machine of the power of up to GW scale, which may release the energy of ~ $10^{12}$ J, with the most of work being spent on rotating the air. Thus, formation of waterspout needs satisfying simultaneously the following two conditions: (i) the presence of vertically aligned cylinder, composed of skeletal-forming dust, floating on the ocean surface, and (ii) the charged cloud above it with high electric potential. These conditions suggest the ways to predicting and preventing the formation of such severe weather phenomena [10].

### 5. CONCLUSIONS

The study of skeletal structures in fusion plasmas, and their interpretation in terms of a fractal condensed matter (FCM) assembled from nanotubular dust, contributes to the rapidly developing studies of dusty component in fusion plasmas (see e.g. [38]) – mostly peripheral one but not only, if the pellet injection is considered – which extend the already established research of dusty plasmas (see e.g. [39,40]) to fusion plasmas.

Probable applications -- both inside and outside the fusion science -- of the FCM which might be responsible for the phenomenon of universal skeletal structures (USS) cover a wide range of phenomena. It is suffice to mention the problem of new nanomaterials (cf. [5(A)]). In the present brief survey, we discussed this problem in applications to severe weather phenomena. For instance, the summary of the problem with respect to tornado is as follows [4(D)].

Tornado, because of its exceptional property of concentrating the energy density in severe weather phenomena, seems to be the best candidate for verifying both the phenomenon of skeletal structuring as itself and the hypothesis for the crucial role of a fractal condensed matter (most probably, nanotubular dust) in severe weather phenomena. The evidences for skeletal structures in tornado suggest them to be a carrier/source of major electrodynamical properties of tornado. This implies that the main body of tornado may be interpreted as a special type of atmospheric dusty plasma. The above approach suggests the probable directions of (i) modeling, in a laboratory electric discharge, the suggested electrodynamical properties of tornado, and (ii) elaborating the technique for early diagnostics of tornado and other severe weather phenomena. The respective laboratory experiments have to model the experimental conditions favorable for electric breakdown in the presence of nanodust.

**Acknowledgments**

We thank our colleagues collaborating with us at this stage of research, especially B.N. Kolbasov and P.V. Romanov. We thank V.I. Kogan for his interest and support.

Partial financial support to our research from the Russian Federation Agency for Atomic Energy and the Russian Foundation for Basic Research (RFBR) is acknowledged. The participation of A.B.K. in the 6-th Symposium was in parts supported by the International Atomic Energy Agency and the RFBR (travel grant RFBR 05-02-26513). The post-Symposium numerical studies of skeletal structuring [34] are supported by the RFBR (project No. 05-08-65507).